# Enhanced magnetic and thermoelectric properties in epitaxial polycrystalline SrRuO$_3$ thin film

*Sungmin Woo[1,2], Sang A Lee[1], Hyeona Mun[3], Young Gwan Choi[4], Chan June Zhung[4], Soohyeon Shin[1], Morgane Lacotte[5], Adrian David[5], Wilfrid Prellier[5], Tuson Park[1], Won Nam Kang[1], Jong Seok Lee[4], Sung Wng Kim[3], and Woo Seok Choi[1]\**

[1]Department of Physics, Sungkyunkwan University, Suwon 16419, Korea
[2]Center for Integrated Nanostructure Physics, Institute for Basic Science, Suwon 16419, Korea
[3]Department of Energy Sciences, Sungkyunkwan University, Suwon 16419, Korea
[4]Department of Physics and Photon Science, Gwangju Institute of Science and Technology, Gwangju 61005, Korea
[5]Laboratorie CRISMAT, CNRS UMR 6508, ENSICAEN, Normandie Universite, F-14050 Caen Cedex 4, France

ABSTRACT

Transition metal oxides thin films show versatile electrical, magnetic, and thermal properties which can be tailored by deliberately introducing macroscopic grain boundaries via polycrystalline solids. In this study, we focus on the modification of the magnetic and thermal transport properties by fabricating single- and polycrystalline epitaxial SrRuO3 thin films using pulsed laser epitaxy. Using epitaxial stabilization technique with atomically flat polycrystalline SrTiO3 substrate, epitaxial polycrystalline SrRuO3 thin film with crystalline quality of each grain comparable to that of single-crystalline counterpart is realized. In particular, alleviated compressive strain near the grain boundaries due to coalescence is evidenced structurally, which induced enhancement of ferromagnetic ordering of the polycrystalline epitaxial thin film. The structural variations associated with the grain boundaries further reduce the thermal conductivity without deteriorating the electronic transport, and lead to enhanced thermoelectric efficiency in the epitaxial polycrystalline thin films, compared with their single-crystalline counterpart.

Grain boundaries in crystals induce non-trivial strain states, leading to phenomena such as dislocations and coalescence. When such long range translational symmetry breaking phenomena occurs, various physical characteristics can be altered, including mechanical, optical, thermal, electric, magnetic, and energy properties. For example, the thermal resistance at the grain boundaries can reduce the thermal conductivity,[1,2] and the strain gradient due to coalescence of crystallites near the grain boundaries can alter the electric and magnetic behaviour of a crystal.[3,4] In addition, the crystallites near the grain boundaries can be modified depending on the size of the grains.[5,6] The simplest way to understand how grain boundaries affect the physical behaviours mentioned above will be to compare single- and polycrystalline solids. In bulk crystals, this comparison is rather straightforward. However, the approach is much less realistic in the case of thin films, because polycrystalline thin film phases usually form under non-optimal growth conditions. Such polycrystalline thin films tend to have inferior crystalline quality with higher defect concentrations and higher chances of being nonstoichiometric compared to the single-crystalline thin films grown under optimal conditions. Accordingly, the lack of high-quality polycrystalline thin films precludes a genuine understanding of the precise role of the grain boundaries.

SrRuO$_3$ (SRO) is a suitable model system for comparing various physical properties of single- and polycrystalline thin films. Perovskite SRO is an itinerant ferromagnet with *p*-type conduction and a Curie temperature ($T_C$) of ~160 K, at which its magnetic properties are closely coupled to its transport and thermal properties.[7,8] Below $T_C$, the electric resistivity follows a Fermi liquid (FL)-associated temperature dependence, which starts with a kink at $T_C$. The fundamental physical properties of SRO thin film can also be tuned by applying epitaxial strain or lattice distortion. For example, pressurizing along different crystallographic orientations of epitaxial SRO thin films using various substrate orientations resulted in significant changes in their magnetic ground states.[9-11] Also, a recent theoretical study predicted that the thermopower in SRO can be largely tuned by the octahedral tilt and rotation, indicating a strong interaction between structural, electric, magnetic, and thermal properties of the complex oxide.[12]

In this study, we investigated the magnetic and thermal transport properties of epitaxial single- and polycrystalline SRO thin films in the context of long range translational symmetry breaking. To minimize changes in other variables except for the existence of grain boundaries (*e.g.*, stoichiometry, crystallinity of each grain, and film thickness), we employed epitaxial stabilization technique using polycrystalline SrTiO$_3$ (poly-STO) substrates to grow polycrystalline SrRuO$_3$ (poly-SRO) thin films. Using atomically flat poly-STO as the growth template, epitaxial poly-SRO thin films could be fabricated under the same ideal growth conditions as the high-quality single-crystalline thin films.[13] Figs. 1(a) and 1(b) schematically show the epitaxial polycrystalline thin films (Fig. 1(b)) in comparison to the conventional epitaxial single-crystalline thin films (Fig. 1(a)). Each domain (grain) within the polycrystalline substrate promotes epitaxial growth of a perovskite thin film domain with the same crystallographic orientation, leading to an epitaxial polycrystalline thin film. The growth of the epitaxial polycrystalline thin films thus enables a reliable comparison of the properties of single- and polycrystalline thin films.

**Results and discussion**

Bulk poly-STO substrates were fabricated by spark plasma sintering, and atomically smooth surfaces were obtained using a chemical and thermal process described elsewhere.[13] Conventional single-crystalline STO (single-STO) substrates with (100), (110), and (111) surface orientations (selected to probe different crystallographic orientations of the thin films), as well as poly-STO were used as substrates for the pulsed laser epitaxy (PLE) of epitaxial SRO thin films. The polycrystalline template enabled the formation of epitaxial poly-SRO thin films at growth conditions identical to those used for the synthesis of high-quality single-crystalline SRO (single-SRO) thin films.[13]

The x-ray diffraction (XRD) results confirm the successful fabrication of epitaxial single- and poly-SRO thin films. Figs. 1(c) and 1(d) show XRD $\theta$-$2\theta$ scans of epitaxial single- and poly-SRO thin films, respectively. The single-SRO thin films exhibit sharp diffraction peaks and clear Pendellösung fringes, which indicate the growth of high-quality epitaxial thin films. According to the XRD reciprocal space maps (shown in Fig. S1, ESI†), the single-SRO thin films (thickness of ~30nm) are fully strained without any lattice relaxation. On the other hand, the poly-SRO thin films exhibit a limited number of weak SRO film peaks, in comparison with the peaks corresponding to the poly-STO substrate, as shown in the bottom panels of Fig. 1(d). The weak intensity can be understood in terms of the x-ray beam cross-section during the XRD scans. Because the average grain size of the polycrystalline film is approximately 10 µm, the cross-section decreases by about four orders of magnitude for these samples. The exact epitaxial relationship between the polycrystalline thin film and the substrate can be obtained from electron beam scattering (EBSD) inverse pole figure and atomic force microscopy (AFM). From topographic and inverse pole figure images in Figs. 1(e), 1(f) and Fig. S2, ESI†, the grains in the poly-SRO thin film are grown with the same crystallographic orientation as the corresponding grains of the poly-STO substrate. Such epitaxial growth has also been suggested by several previous studies. For example, EBSD have been used to determine the epitaxial relationship in thick polycrystalline $Fe_2O_3$ and $BiFeO_3$ films grown on substrates such as $SrTiO_3$, $TiO_2$, and $LaAlO_3$.[14-16]

The temperature-dependent resistivity ($\rho(T)$) and magnetization ($M(T)$) measurements, shown in Figs. 2(a) and 2(b), respectively highlights the ferromagnetic behaviour of the SRO thin films and the effect of the structural modulation. A sudden change in the slope of the $\rho(T)$ curve at $T \sim 150$ K, suggests the increased mean free path of the charge carriers with spin ordering below $T_C$.[17,18] The $T_C$ values can also be estimated from the onset of $M(T)$ below ~150K. The results are summarised in Fig. 2(c). Whereas the small deviation (within 3 K) between the $T_C$ values extracted from the $M(T)$ and $\rho(T)$ data might have various causes, including the different approach used for determining $T_C$ and/or experimental errors, the ferromagnetic $T_C$ systematically increased in the following order for both types of the measurements: (100) < (110) < (111) oriented single- < poly-SRO. Surprisingly, the highest $T_C$ was obtained for the epitaxial poly-SRO thin film. We further analysed the electric and magnetic behaviour in SRO thin films using the relation $\rho(T) = \rho_0 + AT^\alpha$ (where $\rho_0$ is residual resistivity, A is a coefficient, and $\alpha$ is a scaling parameter) in three different $T$ regions (Fig. S3, ESI†). SRO is known to exhibit a bad metallic behaviour for $T > T_C$ ($\alpha =$ 0.5), a crossover regime with substantial scattering effect from the localised electrons for $50 < T < 120$ K ($\alpha = 1.5$), and FL behaviour for $T < 30$ K ($\alpha = 2.0$).[19] The trend observed for $\rho(T)$ of the single-SRO thin films was consistent with the literature, *i.e.*, $\rho_0$ was decreasing in the following sequence: (100) > (110) > (111).[20,21] Furthermore, the smallest $\rho_0$ value was found for the poly-SRO sample, indicating that the crystalline disorder does not significantly influence the electronic transport behaviour in poly-SRO thin films. This leads to the conclusion that the grain boundaries are conducting, and do not substantially hamper the electronic transport across the boundaries. This further suggests that the grain size dependence would be rather small at least in the electronic transport point of view.

The observed variation in the $T_C$ of the single-SRO epitaxial thin films can be understood in terms of the different atomic distortions induced by the different surface orientations. In particular, the (100), (110), and (111) surface orientations of epitaxial thin films are ideally expected to undergo tetragonal, monoclinic, and trigonal distortions, respectively,[22] while the actual structure includes rotation and tilting of the $RuO_6$ octahedra. One of the main consequences of the different surface symmetry is the change in unit cell volume (and in the resulting distance between the neighbouring Ru atoms), as shown in Fig. 3 and Fig. S4, ESI† for the epitaxial SRO thin films. The filled symbols represent the present experimental results out of more than 20 thin film samples. Whereas small deviations within the data corresponding to the same surface orientations are due to intentional modifications in the growth conditions, *e.g.*, in the oxygen partial pressure during the PLE growth, the unit cell volume shows a clear decrease when the surface orientation changes from (100) to (110) to (111).[9-11,20]

The above observations show that $T_C$ scales remarkably well with both unit cell volume and average atomic distance. As the unit cell volume decreases, the $T_C$ systematically decreases, as shown in Fig. 3(a). This result is also consistent with previous studies. In the case of bulk SRO (indicated by empty star symbols in the Fig. 3), a large increase in the lattice parameters and a corresponding decrease in $T_C$ can be achieved by introducing Ru and O vacancies.[23] The highest $T_C$ value was obtained when the unit cell volume approached the bulk value of bulk SRO. In the case of the thin films (indicated by empty symbols, all grown on STO substrates), a further decrease in the unit cell volume was achieved via biaxial compressive strain, and $T_C$ values comparable with those of stoichiometric bulk SRO could be obtained.[20,24] The systematic change in the unit cell volume can be interpreted in a similar way to the change in the average atomic distance (Fig. 3(b), Fig. S4, ESI†).[20] In general, the ferromagnetic $T_C$ of SRO is determined by the exchange interaction within the Ru-O-Ru complex, which can be adjusted by varying interatomic distances and bond angles.[25] In particular, the different substrate symmetry induced distortion in the single-crystalline (100), (110), (111), and polycrystalline film result in different exchange interaction which modifies the $T_C$.

Unfortunately, the above interpretation of the ferromagnetic $T_C$ changes in single-SRO thin films could not be directly applied to the epitaxial poly-SRO thin film, which exhibits the highest $T_C$. First, since the epitaxial poly-SRO thin film exhibits many different crystallographic orientations (surface symmetries), it would be unrealistic to consider a specific atomic distortion based on symmetry considerations. Second, it was rather difficult to extract the exact lattice parameters of the poly-SRO thin films experimentally, owing to the weak XRD intensity of the relevant peaks. Nevertheless, from the single-SRO analyses, it is evident that the enhancement in the $T_C$ of poly-SRO thin film originates from the modifications in the atomic structure. As an alternative interpretation, we note that subtle change in the epitaxial strain can modify the $T_C$ of poly-SRO thin film. For examples, SRO thin films (with a pseudo-cubic lattice constant of 3.93 Å) grown on GdScO$_3$ substrates (with a pseudo-cubic lattice constant of 3.96 Å) instead of STO substrates (with a cubic lattice constant of 3.91 Å) exhibit a larger $T_C$ compared to bulk single-crystal SRO.[26] The biaxial tensile strain, which is known to primarily enlarge the in-plane bond angle, leads to an increased width of the Ru 4$d$ electron band. The resulting increase in the exchange energy could lead to the observed enhancement of the ferromagnetic $T_C$.

We note that coalescence of the grains in the poly-SRO thin film can alleviate the original epitaxial compressive strain between the SRO thin film and the STO substrate, which could facilitate magnetic ordering. Coalescence occurs when two or more grains merge together during the growth of thin films.[3,27,28] Fig. 4(a) schematically shows the possible coalescence behaviour during the growth of epitaxial poly-SRO thin films. In the initial stage, the SRO nucleates on the individual STO grains. Owing to the lattice mismatch between film and substrate, each grain experiences conventional epitaxial compressive strain. As the growth proceeds, the grains approach each other and start to merge, relieving the compressive strain. Consequently, the epitaxial poly-SRO thin film can be expected to experience a lower degree of compressive strain than the single-SRO thin films.

Experimental evidence of the alleviated strain state in epitaxial poly-SRO thin films can be obtained from the topography images and thickness dependent XRD scans. Fig. 4(b) shows the topographic images of the poly-STO substrate and poly-SRO thin films obtained by atomic force microscopy (AFM). The poly-STO substrate (left image) typically exhibits valleys with a depth of ~30 nm at the grain boundaries. As such gaps are not observed for the corresponding thin film (right image), coalescence near the grain boundaries would have taken place at high growth temperatures. Additional coalescence between the nanoscale-sized (~100 nm) islands on top of each SRO grain might also be expected based on the AFM images of the thin film. Moreover, the overall reduction in the compressive strain in the poly-SRO thin film can be demonstrated by the XRD $\theta$-$2\theta$ scans. As shown in Fig. S5, ESI†, the out-of-plane lattice parameters for SRO grains with (100) and (110) orientations within the poly-SRO thin film are slightly smaller than those of the (100) and (110) oriented single-SRO thin films, respectively. The smaller out-of-plane lattice parameters observed in the poly-SRO thin films indirectly suggest a different strain state (less compressive strain), even though the single- and poly-SRO films were grown under the same conditions. The thick poly-SRO thin film shows even smaller lattice parameters, indicating that the compressive strain is relieved with increasing film thickness, as shown in Fig. 4(a). To conclude, the enhanced ferromagnetism indicates the unexpected strain state in the poly-SRO thin films, which also influences the thermal transport properties as follows.

Fig. 5 shows the Seebeck coefficient ($S$) along with the electric and thermal conductivity ($\kappa$) of the epitaxial SRO thin films. We specifically note that such thermal transport measurements on thin films are largely limited, in part because it is difficult to exclude the effect of the substrate in $\kappa$ measurements. While $\kappa$ of bulk SRO has been frequently reported,[29-32] measurement of $\kappa$ of the thin film SRO has not yet been reported. As shown in Fig. 5(a), the temperature-dependent Seebeck coefficient ($S(T)$) is almost unaffected by the temperature and crystallographic orientation. The $S(T)$ values (~37 µV K$^{-1}$ on average) were similar to those measured for bulk SRO,[31] which suggests a *p*-type conduction behaviour as expected. This result indicates that the macroscopic thermal diffusion of charge carriers is not significantly influenced by grain boundaries and/or crystallographic orientations. On the other hand, the temperature-dependent electric conductivity ($\sigma(T)$) shows a systematic dependence on the surface orientations. $\sigma(T)$ decreases as the orientation changes from (100) to (110) to (111) in single-SRO thin films, and the lowest value is obtained for the epitaxial poly-SRO thin film. However, the reduction of $\sigma(T)$ in the epitaxial poly-SRO thin film is not substantial, suggesting that the existence of various surface orientations, instead of the grain boundaries, might be sufficient to explain the observed differences. This observation is consistent with our previous analyses of $\rho(T)$ at low temperatures. The results obtained for both $S(T)$ and $\sigma(T)$ indicate that the effect of grain boundaries on the electrical transport is relatively minor, and that the grain boundaries in poly-SRO thin films can be electrically considered as Ohmic junctions, as discussed previously.

On the other hand, the grain boundaries were found to reduce $\kappa$ significantly in the poly-SRO thin film. The $\kappa$ values of epitaxial single- and poly-SRO thin films were determined using time domain thermal reflectance (TDTR) measurements at room temperature. The results are shown in Fig. 5(b). As for $S(T)$, we did not observe large differences in the $\kappa$ values among the single-SRO thin films. The obtained values (~8.3 W m$^{-1}$ K$^{-1}$ on average) were slightly larger than those of bulk SRO (~6 W m$^{-1}$ K$^{-1}$).[33,34] However, $\kappa$ of the poly-SRO thin film was determined to be ~5.6 W m$^{-1}$ K$^{-1}$ on average, which is about ~32% lower than the average value measured for the single-SRO thin films. The lower $\kappa$ measured for the poly-SRO thin film can be attributed to addition of thermal resistance near the grain boundaries. The phonon mean free path of the SRO is ~15 nm, which is much smaller than the grain size of the epitaxial poly-SRO thin film (~10 μm).[35,36] In such case, scattering at the grain boundaries can be neglected and the reduction of $\kappa$ is limited. For this reason, nanosized grains with size comparable to the phonon mean free path can generally be used to reduce $\kappa$.[37-41] On the other hand, in such nanosized grains, the long-range translational symmetry is broken, usually accompanied by a drop in $\sigma$ and $S$.[42-44] In the present case, relatively large grains are connected by thick, metallic boundaries with strain gradient, which results in a significant decrease in $\kappa$, while maintaining $\sigma$ unchanged.

Furthermore, the reduction of $\kappa$ mainly originates from the decrease in phonon (lattice) thermal conductivity ($\kappa_p$), supporting the above conclusion. The thermal conductivity is composed of the electrical thermal conductivity ($\kappa_e$) and the phonon thermal conductivity ($\kappa = \kappa_e + \kappa_p$). In general, changes in crystallinity, defect concentration, and crystallographic anisotropy in crystalline thin films can affect both $\kappa_e$ and $\kappa_p$. However, the $\kappa_e$ values of epitaxial single- and poly-SRO thin films were similar, as shown in Fig. 5(b). For the single-SRO thin film, $\kappa_e$ was ~1.9 W m$^{-1}$ K$^{-1}$, comprising ~21% of the total $\kappa$. This fraction increased up to 27% for the poly-SRO thin film ($\kappa_e$ = ~1.5 W m$^{-1}$ K$^{-1}$), suggesting a large drop in $\kappa_p$ from 6.3-6.7 W m$^{-1}$ K$^{-1}$ for single-SRO to ~4.1 W m$^{-1}$ K$^{-1}$ for poly-SRO. The marked decrease in $\kappa_p$ in the poly-SRO thin film can be attributed to the interfacial thermal resistance at the grain boundaries. These structural defects limit phonon propagation.[45-47] We also measured $\kappa$ of the single- and poly-STO substrates to further confirm that the reduction in $\kappa_p$ is indeed an effect of the grain boundaries. While the relative reduction is smaller in this case (−26% for STO compared with −32% for SRO), possibly due to the absence of additional strain effects in the bulk crystal, the significant reduction in $\kappa_\text{p}$ for bulk poly-STO clearly demonstrates the importance of the grain boundaries also in the case of microsized grains.

Using the above $\kappa$ measurements for epitaxial single- and polycrystalline thin films, an enhanced thermoelectric figure of merit (*ZT*) was estimated, as shown in Fig. 5(c). The *ZT* values were calculated as, $ZT = (\sigma S^2 T)/\kappa$. The average *ZT* value obtained for single-SRO ((100), (110), and (111)) thin films and for the epitaxial poly-SRO thin film were ~0.0145 and ~0.0173, respectively. While the *ZT* value of the poly-SRO thin film is not particularly high, even among *p*-type oxide semiconductors, the 19% increase in *ZT* at room temperature highlights the effectiveness of the present materials design to reduce $\kappa$ via epitaxial polycrystalline thin films for thermoelectric applications.

**Conclusions**

We investigated the magnetic, electric, and thermal properties of epitaxial polycrystalline SrRuO$_3$ thin films in comparison with single-crystalline SrRuO$_3$ thin films. Enhanced ferromagnetic $T_\text{C}$ was observed in epitaxial polycrystalline SrRuO$_3$ thin films, which was attributed to strain gradient near the grain boundaries. The thermal resistance in grain boundaries of the polycrystalline thin film also resulted in a reduced lattice thermal conductivity, which in turn led to a higher *ZT* value. The present approach provides a new route for tailoring magnetic and thermal properties via long-range crystalline symmetry engineering in epitaxial thin films.

**Experimental**

**Epitaxial thin film growth and structural characterization**

High-quality epitaxial single- and polycrystalline SrRuO$_3$ thin films were grown on atomically flat SrTiO$_3$ single- and polycrystalline substrates using PLE at 700 °C. A 248 nm laser (248 nm; IPEX 864, Lightmachinery) with fluence of 1.5 J/cm$^2$ and repetition rate of 2 Hz was used. SrRuO$_3$ thin films were grown under 30 and 100 mTorr oxygen partial pressures.[48,49] We have fabricated more than six sets (one set comprising of single-crystalline (100), (110), (111), and polycrystalline SrRuO$_3$) of the epitaxial thin films, which showed essentially the same results. Each set of thin films was fabricated at the same time of the growth. The thickness of the single-crystalline SrRuO$_3$ thin films was 30 ± 1 nm, as measured by x-ray reflectometry (XRR). The atomic structure, crystal orientation, and epitaxy relationship of the SrRuO$_3$ thin films were characterised using high-resolution x-ray diffraction (XRD) (Rigaku, Smartlab). Electron backscattering diffraction (EBSD), JSM7000F. For the EBSD, the sample was mounted at a 70° angle from scanning electron microscope operated at 20 kV. The probe current of the aperture was 1×10$^{-8}$ A and the Kikuchi diagrams were recorded with the beam step size of 0.5 μm. For image processing, a commercial program (OIM Analysis 5.31) was used.

**Electrical transport and Seebeck coefficient measurements**

The temperature-dependent resistivity, $\rho(T)$, was measured using a low-temperature closed-cycle refrigerator (CCR). The measurements were performed from 300 to 20 K, using the Van der Pauw method with in electrodes and Au wires. The Seebeck coefficient and electrical conductivity in the high-temperature region (300 to 750 K) were measured with a ZEM-3 (ULVAC-RIKO ZEM-3) system.

**Magnetic moment measurements**

The temperature-dependent magnetization, $M(T)$, was measured between 2 and 300 K at a magnetic field of 100 Oe, using a Magnetic Property Measurement System (MPMS, Quantum Design). An in-plane geometry was used to measure the magnetization of the thin films.

**Thermal conductivity measurements**

The thermal conductivity was measured by the time-domain thermoreflectance (TDTR) technique.[50] Femtosecond laser pulses of 800 nm wavelength were focused on a 100 nm-thick Al transducer that was evaporated on the $SrRuO_3$ thin films, and the thermoreflectance was monitored by probe laser pulses with a time delay of up to 4 ns. The $SrRuO_3$ thermal conductivity and the Al/$SrRuO_3$ thermal boundary conductance were obtained by fitting the obtained thermoreflectance curve using the parameters below, taking into account the heat conduction from the Al layer to the $SrRuO_3$ film and then to the $SrTiO_3$ substrate. An Al thermal conductivity of 127 W m$^{-1}$ K$^{-1}$ was obtained by applying the Wiedemann-Franz law to the measured electric conductivity. The thermal conductivities of the single-crystalline and polycrystalline $SrTiO_3$ substrates, obtained from separate TDTR measurements, were 10.20 and 7.63 W m$^{-1}$ K$^{-1}$, respectively. Heat capacity values of 2.86 J cm$^{-3}$ K$^{-1}$ for $SrRuO_3$[31] and 2.74 J cm$^{-3}$ K$^{-1}$ for $SrTiO_3$ were obtained from the literature.[51,52] Because the thermoreflectance response has a low sensitivity to the $SrRuO_3$/$SrTiO_3$ boundary conductance, we did not include it as a variable in the fitting process, but used the literature value of 700 MW m$^{-2}$ K$^{-1}$ instead.[53] The electrical thermal conductivity can be calculated as $\kappa_e = L\sigma T$, where $L$ is the Lorenz number (2.09×10$^{-8}$ W Ω$^{-1}$ K$^{-2}$),[29] $\sigma$ is the electric conductivity, and $T$ is the absolute temperature.

**Authorship contributions**

S.M. and S.A.L. fabricated and characterized the thin films; H.M. and S.W.K. measured the Seebeck coefficients; Y.G.C., C.J.Z, and J.S.L. measured and analysed the thermal conductivities; S.S., T.P., and W.N.K. performed the magnetization measurements; M.L., A.D., and W.P. fabricated the polycrystalline substrates; W.S.C. conceived and led the study; S.M. and W.S.C. wrote the manuscript with the help of all the other authors.

**Acknowledgements**

We appreciate valuable discussions with C.U. Jung and T.S. Yoo. This research was supported by the International Research & Development Program of the National Research Foundation (NRF) of Korea funded by the Ministry of Science, ICT, and Future Planning of Korea (2016K1A3A1A21004685, 2012R1A3A2048816 (SS & TP) and 2015R1A1A1A05001560 (YGC, CJZ & JSL)) and the Institute for Basic Science (IBS-R011-D1) and Programme Asie 127096 supported by the Conseil Regional Normandie, and PHC Star 36453VL (ML, AD & WP).
15

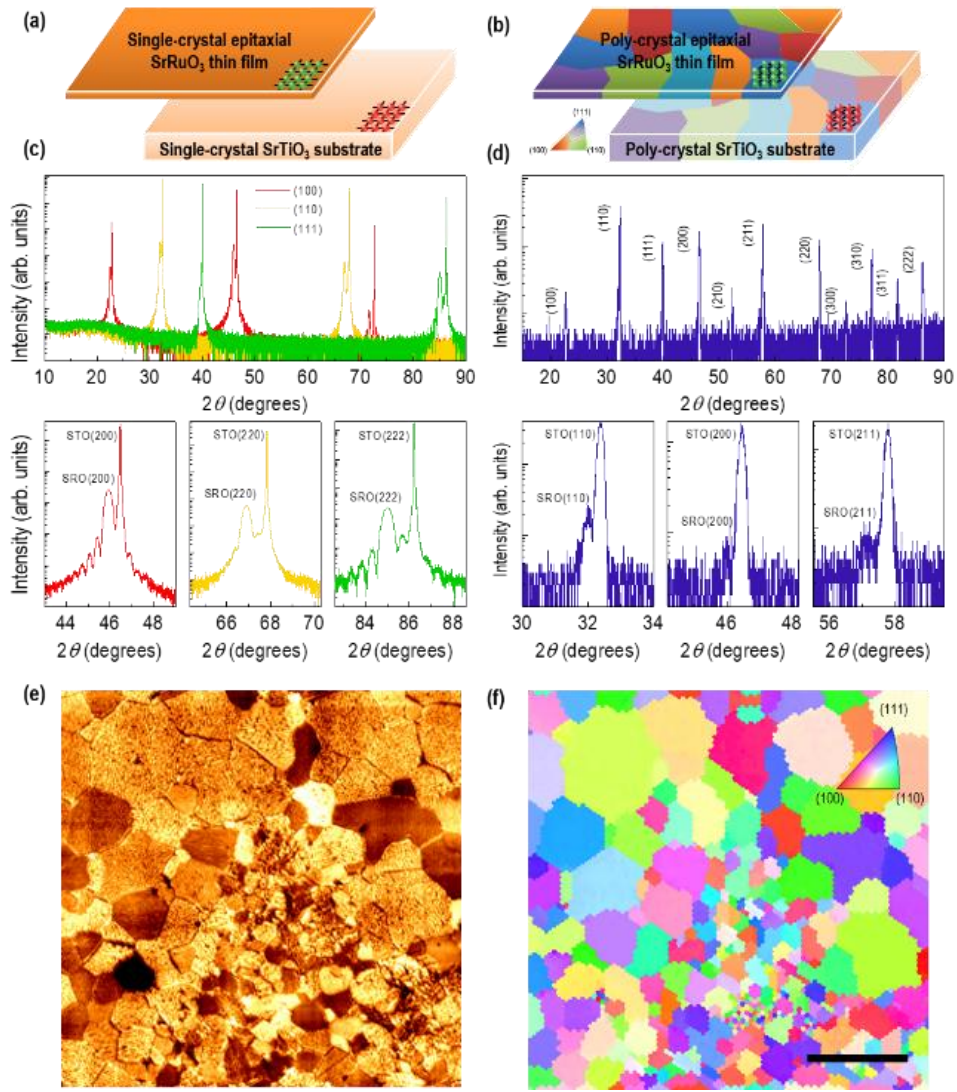

**Figure 1. Schematic representations and crystal structures of single- and polycrystalline epitaxial thin films.** Schematic diagrams of (a) single- and (b) polycrystalline epitaxial thin films fabricated on single- and polycrystalline substrates, respectively. High-resolution XRD $\theta$-$2\theta$ scans for (c) single- and (d) polycrystalline epitaxial SrRuO$_3$ thin films fabricated on SrTiO$_3$ substrates. Single-crystalline thin films are deposited on (100) (red), (110) (yellow), and (111) (green) SrTiO$_3$ substrates. The lower panels show magnified views of the regions near the main diffraction peaks

of the SrTiO$_3$ substrates. (e) AFM topographic and (f) EBSD inverse pole figure images of the epitaxial polycrystalline SrRuO$_3$ thin film. The scale bar represents 10 μm.

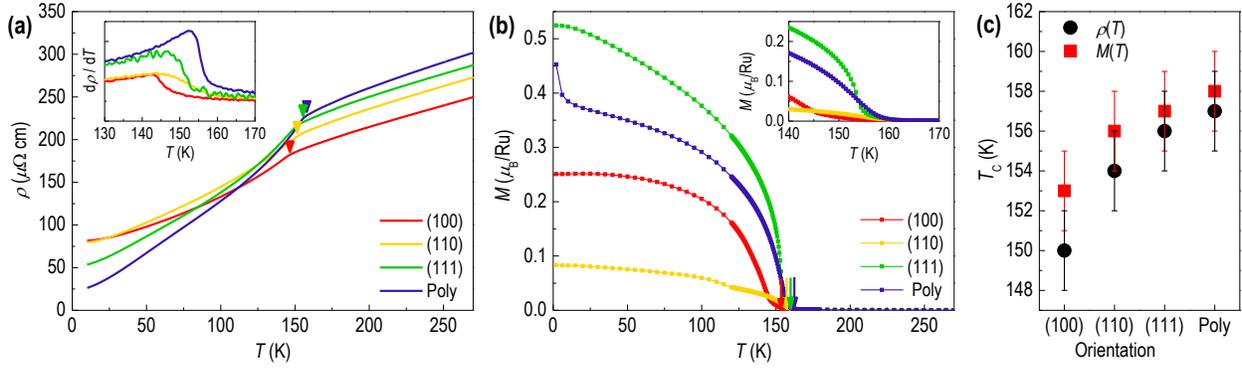

**Figure 2. Electronic and magnetic responses.** (a) $\rho(T)$ and (b) $M(T)$ for single- and polycrystalline SrRuO$_3$ epitaxial thin films deposited on single-crystalline (100) (red), (110) (yellow), (111) (green), and polycrystalline (navy) SrTiO$_3$ substrates. The arrows indicate the $T_C$ values. The insets of panels (a) and (b) show d$\rho(T)$/d$T$ and $M(T)$ near $T_C$. (c) $T_C$ dependence on epitaxial thin film orientation. The black circles and red squares denote the $T_C$ values obtained from $\rho(T)$ and $M(T)$ measurements, respectively.

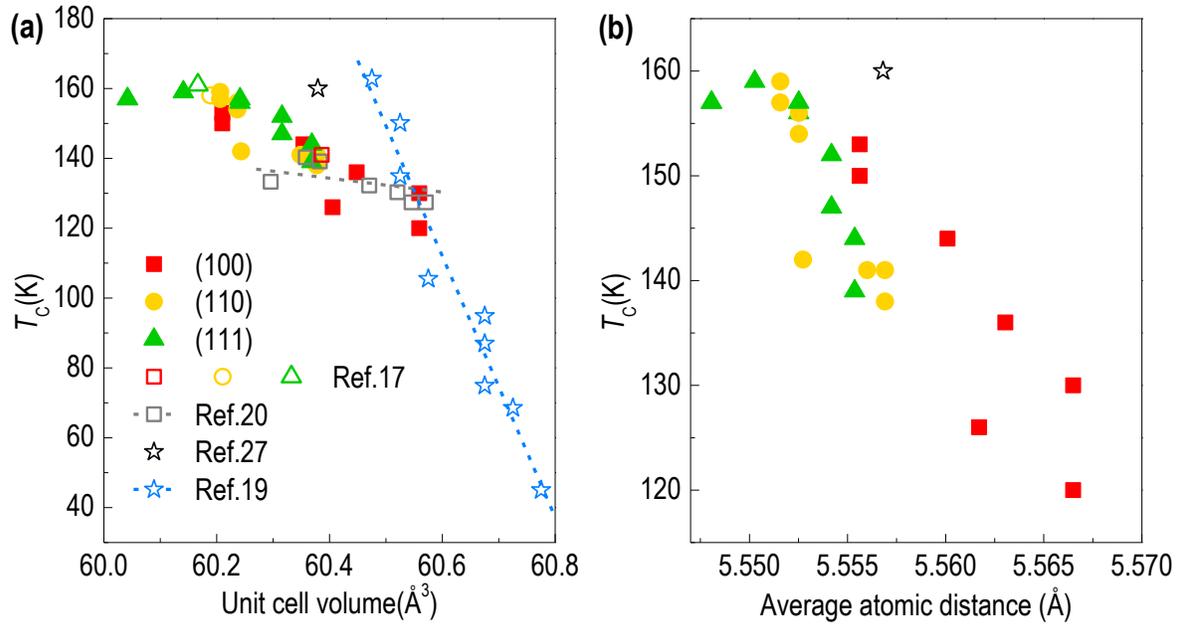

**Figure 3. Scaling of ferromagnetic ordering in SrRuO$_3$ epitaxial thin films.** (a) $T_C$ plotted as a function of the unit cell volume. For comparison, the $T_C$ data taken from Refs. 20, 22, 23, and 30 are also shown as empty symbols. (b) $T_C$ plotted as a function of average atomic distance.

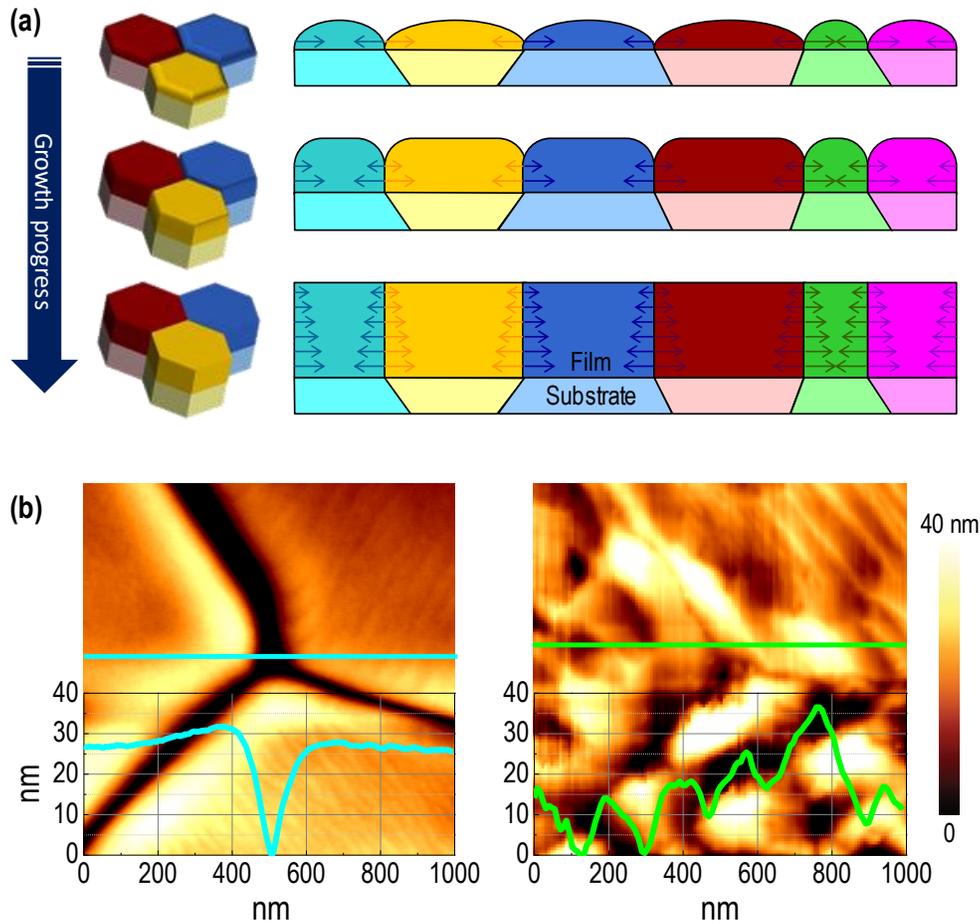

**Figure 4. Coalescence in epitaxial polycrystalline thin films.** (a) Schematic diagram of the growth of the epitaxial polycrystalline thin film. The right panels show the cross-section of the thin film during the growth process. During the initial growth stages, the thin film is subject to epitaxial compressive strain imposed by the lattice mismatch between the substrate and the thin film. As the thin film grows further, coalescence between the islands in each grain generates additional tensile strain, which relieves the original compressive strain. (b) AFM topographic images of polycrystalline $SrTiO_3$ substrate (left) and $SrRuO_3$ epitaxial thin film (right) near a grain boundary.

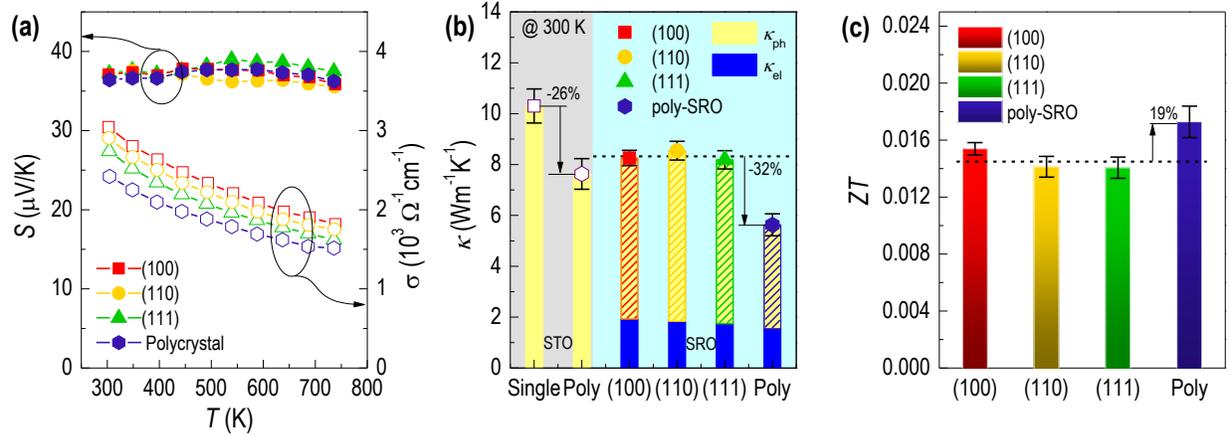

**Figure 3. Thermal transport properties.** (a) Temperature-dependent Seebeck coefficient ($S$) and electric conductivity ($\sigma$) of single-crystalline SrRuO$_3$ (100) (red squares), (110) (yellow circles), (111) (green triangles), and polycrystalline SrRuO$_3$ thin films (navy hexagons). $S(T)$ is approximately constant and thus independent of the crystalline orientations, whereas $\sigma(T)$ shows some changes, consistent with Fig. 2(b). (b) Electrical ($\kappa_e$) and phonon ($\kappa_p$) thermal conductivities of single- and polycrystalline SrRuO$_3$ epitaxial thin films. The dotted line represents the average $\kappa$ values of the single-crystalline thin films. (c) Thermoelectric figure of merit ($ZT$) of single-crystalline (100), (110), (111), and polycrystalline SrRuO$_3$ epitaxial thin films. Dotted lines denote the average $ZT$ of single-crystalline SrRuO$_3$ thin films.